# Long-standing problem: The nuclear level density angular-momentum dependence and isomeric data assessment

M. Avrigeanu,[1,*] E. Šimečková,[2,†] J. Mrázek,[2] X. Ledoux,[3] J. Novak,[2] M. Štefánik,[2] M. Ansorge,[2] A. Cassisa,[2] J. Kozic,[2] C. Costache,[1] and V. Avrigeanu[1]

[1]*Horia Hulubei National Institute for Physics and Nuclear Engineering (IFIN-HH), 077125 Bucharest-Magurele, Romania*
[2]*Nuclear Physics Institute (NPI) CAS, 25068 Řež, Czech Republic*
[3]*Grand Accélérateur National d'Ions Lourds, CEA/DRF—CNRS/IN2P3 (GANIL), B.P. 55027, Caen F-14076, France*

Recent $^{91,92,93}$Tc activation for deuterons incident on $^{nat}$Mo has become a challenge for the nuclear level density (NLD) angular-momentum dependence. Actually, replacement of the moment of inertia rigid-body value $I_r$ by half of it, within a given NLD parameter set, demands a change of the rest of NLD parameters significantly beyond their fitted limits. The corresponding uncertainty of calculated cross sections versus the NLD parameter accuracy is also higher, while use of either the same or distinct compound-nucleus and preequilibrium emission spin distributions becomes significant at higher incident energies. Nevertheless, the current way to describe experimental isomeric cross sections by using at most half of $I_r$ values provides agreement of the measured and calculated data at the price of less and less correct NLDs. The moment of inertia relevance for the NLD correctness also emphasizes the value of a direct method to endorse it. Further measurements of average resonance spacings of $s$-wave neutrons and protons, corresponding to different spins of the same nucleus, are therefore highly demanded.

DOI: 10.1103/cjqq-6psq

## I. INTRODUCTION

Recent accurate isomeric cross sections were obtained for deuteron-induced reactions on molybdenum at incident energies up to 40 MeV [1,2]. They become a challenge for the real angular-momentum dependence of both the phenomenological nuclear level density (NLD) and the particle-hole state density, which are central issues of the compound-nucleus (CN) statistical Hauser-Feshbach (HF) [3] and preequilibrium emission (PE) [4] models, respectively. In fact, both densities are most important for basic research as well as various applications from nuclear technology to medical praxis. However, their Gaussian spin dependence, with the square of dispersion known as the spin-cutoff parameter $\sigma^2$, is yet an open question. While $\sigma^2$ is implicit in the Bethe formula [5] based on the Fermi gas (FG) model, assuming equally spaced single-particle states and no collective levels, it was related to rotations of the whole nucleus as a rigid body [6]. Then, Ericson expressed $\sigma^2 = It/\hbar^2$ in terms of the nuclear moment of inertia $I$ and nuclear temperature $t$, with the rigid-body value $I_r$ as a limit at high excitation, where the nucleus can be described as an ordinary FG [7].

Nevertheless, $I$ values reduced by ⩽30% and uncertainty of the same order were found suitable at low excitation due to pairing interactions [7]. Over decades of studies, briefly reviewed elsewhere [8], a recent isomeric data systematics [9] showed that a sound account of measured activation cross sections using the same CN and PE spin distribution corresponds to the ratio $\eta = I/I_r$ value of 0.5. Meanwhile, it was pointed out that a strong decrease of this ratio is artificial and follows an improper use of the CN spin distribution also for the PE modeling [10–13].

On the other hand, the PE spin distribution was early discussed by Feshbach *et al.* [14] and further detailed by Gruppelaar [15] and Fu [16]. However, use of the CN spin distribution also within the PE stage is the default in the worldwide used code talys even though "there has been quite some debate about the correct spin distributions for preequilibrium reactions" [17]. It could be of interest to note that an isomer overproduction was found by similar calculations for proton-induced reactions on $^{93}$Nb at energies from $\sim$20 to 200 MeV [18].

Noticeable overestimation was also found for the deuteron-induced reactions on Mo at incident energies above $\sim$25 MeV [1], corresponding to the talys default option of the same CN and PE spin distribution but $\eta$ values between 0.25 and 1. Smaller effects were found corresponding to distinct PE spin distributions, based on particle-hole state densities. Nevertheless, the impact on the NLD correctness if only the spin-cutoff parameter is changed within an NLD parameter set already fixed by various independent data fit has not yet been considered. The same for use of the CN spin distribution also for the PE account within HF + PE calculations of nucleon- and deuteron-induced reactions as the object of this work.

Various NLD parameter sets obtained by fit of low-lying discrete level numbers and average $s$-wave resonance

---

spacings, for distinct $I$ values between full and half of its $I_r$ value, are considered in Sec. II. Effects of changing only the spin-cutoff parameter within an NLD parameter set, as well as the use of either the same or distinct CN and PE spin distributions in HF + PE model calculations, are discussed in Sec. III for the well-known $21/2^-$ isomer $^{93}$Mo$^m$ within $(n, 2n)$ and $(p, n)$ reactions on $^{94}$Mo and $^{93}$Nb nuclei, respectively, as well as $^{91,92,93,101}$Tc and $^{101}$Mo activation by deuterons incident on $^{nat}$Mo. Conclusions of this work are given in Sec. IV, while our own data used in this paper and preliminary results were presented elsewhere [19].

## II. LEVEL-DENSITY PARAMETER UNCERTAINTIES

### A. Moment of inertia assessment

Fit of low-lying levels numbers $N_d$ and average resonance spacings $D_0^{\text{exp}}$ for $s$-wave nucleons, in the energy range $\Delta E$ above the nucleon separation energy $S$, usually provides the NLD parameters. Within the widely used back-shifted Fermi gas (BSFG) model [20,21], they are the level-density parameter $a$, ground-state (g.s.) shift $\Delta$, and an effective moment of inertia $I$ for the spin-cutoff parameter inference. Two distinct parameter sets were provided formerly [20], with $a$ and $\Delta$ values corresponding to either half or the full $I_r$ values, but then only the latter received due consideration [21].

However, a direct determination of the spin-cutoff parameter by Weigmann *et al.* [22] through comparison of neutron and proton resonance data for the same compound nucleus led to a value of $I/I_r = 0.75 \pm 0.06$ at the neutron binding energy for the nucleus $^{51}$V [23]. Unfortunately, the same analysis could not be made for other nuclei due to fewer average resonance spacings for $s$-wave protons, as well as rather large error bars of these data. Further, an energy-dependent moment of inertia was adopted, following also theoretical predictions [24–27], with the $\eta$ value first between 0.5 and 0.75, for excitation energies from the nucleus g.s. to the nucleon binding energy, and then increased to 1 around $E^* = 15$ MeV. Its use within further analyses of nucleon- and $\alpha$-induced reactions led to a suitable account of isomeric cross-section ratios, too [28]. On the other hand, an ultimate conclusion on this point was precluded because even a change of the $\eta$ value between 0.5 and 1 led to results' variation similar to measured-data spread and decay-scheme effects [8,29].

In the meantime, the error bars of both $N_d$ [30] and $D_0^{\text{exp}}$ data [31–34] were also fitted [35], for the abovementioned energy-dependent $I$ [23]. The corresponding limits of the BSFG parameters $a$ and $\Delta$ have been used to illustrate NLD uncertainty bands of calculated cross sections [28,35–38]. However, they were not involved within parameter adjustment in order to fit the measured reaction cross sections [39], but for assessment of the NLD weight to eventual disagreement between model calculations and measured cross sections. A global overview of the BSFG level-density parameter $a$ for the mass region $A \sim 90$ is displayed in Fig. 1, taking into account the half and full $I_r$ values as well as the energy-dependent value. The parameter limits due to the error bars of the fitted $N_d$ and $D_0^{\text{exp}}$ data given in Table I are also shown for the third case, being similar for the parameter sets with constant $I_r/2$ and $I_r$ values. Actually, this table format is similar to

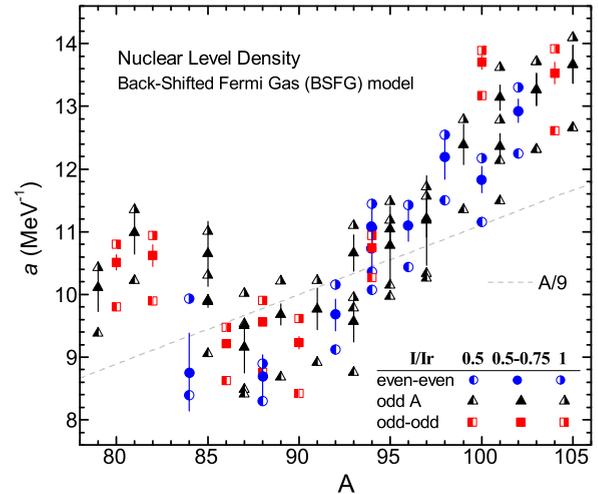

FIG. 1. Mass dependence of BSFG level-density parameter $a$ (Table I) for even-even (circles), odd-$A$ (triangles), and odd-odd (squares) nuclei, with ratios $I/I_r$ of 0.5 (left-half filled symbols), 1 (right-half filled), and between 0.5 and 0.75 from g.s. to $S$ (filled) with error bars related to fitted low-lying discrete levels and average $s$-wave resonance spacings. Usual assumption $A/9$ is also shown (dashed line).

Tables 1 and A1 of Ref. [20], with the additional $a$ and $\Delta$ values corresponding to the energy-dependent $I$ as well as the related parameter limits.

There are thus shown at once uncertainties of the BSFG $a$ and $\Delta$ parameters due to the fitted $N_d$ and $D_0^{\text{exp}}$ data, and the higher variance of these parameters due to the $\eta$-value assumption. Thus, the $I_r$ value with half of it demands actually a change of the corresponding $a$ parameter significantly beyond its limits mentioned above, in order to have a similar level density.

On the other hand, the washing out of shell effects [41,42] has been taken into account above the neutron binding using the method of Koning and Chadwick [43] for fixing the appropriate shell correction energy within an energy-dependent $a$ level-density parameter [28,35–37]. In order to have a smooth connection, a transition range from the BSFG description to the higher-energy approach was chosen between the neutron binding energy and the excitation energy of 15 MeV.

Last but not least, for the nuclei without resonance data the smooth-curve method [44] has been applied by using average $a$ values, for the fit of the low-lying discrete levels. The $A$-range systematics of $a$, for distinct even-odd nucleon numbers in Table I, have been used in this respect. Obviously, the uncertainties of these average $a$ values have often been larger than those for the similar parameters of nuclei with resonance data.

### B. Spin-dependent level densities for distinct $\eta$ values

Distinct spin-dependent level densities are illustrated for each of the three $I/I_r$ ratios in Fig. 2, at the excitation energy of the $s$-wave neutron average resonance spacing of $^{95}$Mo nucleus (Table I). There is also shown the spin dependence corresponding to additional fit of both the lower and higher limits of $D_0^{\text{exp}}$ [31,32], for the energy-dependent moment of

TABLE I. Low-lying levels number $N_d$ up to excitation energy $E_d^*$ [30] and average s-wave resonance spacings $D_0^{\mathrm{exp}}$ ([31] if not otherwise mentioned, and with uncertainties given in units of the last digit in parentheses) in the energy range $\Delta E$ [40] above the separation energy $S$, for the target-nucleus g.s. spin $I_0$, fitted to obtain the BSFG level-density parameter $a$ and g.s. shift $\Delta$, for a spin-cutoff parameter calculated with the moment of inertia $I$ corresponding to either half of the rigid-body value $I_r$, full $I_r$ value, or a change between half and 75% of $I_r$, from g.s. to $S$ [23], besides a reduced radius $r_0 = 1.25$ fm.

| Nucleus | Fitted level and resonance data | | | | | $I = 0.5 I_r$ | | $I = I_r$ | | $I = (0.5\text{–}0.75) I_r$ | |
|---|---|---|---|---|---|---|---|---|---|---|---|
| | $E_d^*$ (MeV) | $N_d$ | $S + \frac{\Delta E}{2}$ (MeV) | $I_0$ | $D_0^{\mathrm{exp}}$ (keV) | $a$ (MeV$^{-1}$) | $\Delta$ (MeV) | $a$ (MeV$^{-1}$) | $\Delta$ (MeV) | $a$ (MeV$^{-1}$) | $\Delta$ (MeV) |
| Even-even | | | | | | | | | | | |
| $^{84}$Kr | 3.476 | 28(2) | 10.52 | 9/2 | 0.2(1) | 8.39 | 0.65 | 9.93 | 0.61 | 8.76(62) | 0.74(16) |
| $^{88}$Sr | 4.801 | 47(2) | 11.114 | 9/2 | 0.29(8) | 8.3 | 1.52 | 8.9 | 1.52 | 8.70(34) | 1.63(11) |
| $^{92}$Zr | 3.725 | 54(2) | 8.647 | 5/2 | 0.55(10) | 9.12 | 0.65 | 10.16 | 0.78 | 9.67(25) | 0.79(6) |
| $^{94}$Zr | 3.156 | 27(2) | 8.218 | 5/2 | 0.302(75) | 10.36 | 0.9 | 11.45 | 1.06 | 11.07(33) | 1.02(3) |
| $^{90}$Mo | 3.185 | 22(2) | | | | 8.55 | 0.55 | 9.5 | 0.63 | 9.15(50) | 0.71(10) |
| $^{92}$Mo | 4.019 | 36(5) | | | | 8.8 | 1.11 | 9.9 | 1.47 | 9.40(33) | 1.28(2) |
| $^{94}$Mo | 3.462 | 61(2) | 9.678 | 5/2 | 0.081(24)[a] | 10.07 | 0.62 | 11.06 | 0.7 | 10.74(40) | 0.78(9) |
| $^{96}$Mo | 2.875 | 38(2) | 9.154 | 5/2 | 0.081(14), 0.0662(30)[b] | 10.44 | 0.42 | 11.43 | 0.47 | 11.10(25) | 0.55(3) |
| $^{98}$Mo | 2.701 | 38(2) | 8.643 | 5/2 | 0.06(1), 0.047(6)[a] | 11.5 | 0.49 | 12.55 | 0.52 | 12.20(36) | 0.60(4) |
| $^{100}$Mo | 2.432 | 31(2) | | | | 11.8 | 0.38 | 12.8 | 0.4 | 12.5(4) | 0.48(4) |
| $^{100}$Ru | 2.915 | 61(2) | 9.674 | 5/2 | 0.025(4) | 11.16 | 0.35 | 12.17 | 0.4 | 11.83(21) | 0.48(3) |
| $^{102}$Ru | 2.467 | 27(2) | 9.22 | 5/2 | 0.018(3) | 12.25 | 0.56 | 13.3 | 0.58 | 12.93(19) | 0.65(1) |
| Odd-odd | | | | | | | | | | | |
| $^{80}$Br | 0.919 | 56(2) | 7.897 | 3/2 | 0.047(5) | 9.8 | −2.03 | 10.8 | −1.92 | 10.51(12) | −1.80(1) |
| $^{82}$Br | 1.276 | 44(2) | 7.601 | 3/2 | 0.100(15) | 9.9 | −1.37 | 10.94 | −1.28 | 10.62(17) | −1.18(2) |
| $^{86}$Rb | 1.559 | 24(2) | 8.661 | 5/2 | 0.172(8) | 8.63 | −1.27 | 9.48 | −1.07 | 9.21(1) | −0.96(6) |
| $^{88}$Rb | 1.442 | 21(2) | 6.094 | 3/2 | 1.63(15) | 8.76 | −1.13 | 9.91 | −0.99 | 9.56(6) | −0.9(5) |
| $^{90}$Y | 2.327 | 29(2) | 6.857 | 1/2 | 3.7(4) | 8.42 | −0.57 | 9.62 | −0.38 | 9.23(10) | −0.32(2) |
| $^{92}$Nb | 1.851 | 41(2) | | | | 9.3 | −1.0 | 10.3 | −0.89 | 10.0(2) | −0.79(3) |
| $^{94}$Nb | 1.281 | 41(2) | 7.232 | 9/2 | 0.094(10) | 10.27 | −1.41 | 10.94 | −1.43 | 10.75(12) | −1.29(1) |
| $^{90}$Tc | 0.340 | 6(2) | | | | 9.5 | −1.61 | 10.2 | −1.62 | 10.0(4) | −1.48(22) |
| $^{92}$Tc | 0.686 | 8(2) | | | | 9.5 | −1.2 | 10.6 | −1.13 | 10.4(4) | −1.00(8) |
| $^{94}$Tc | 1.448 | 26(2) | | | | 10.5 | −0.78 | 11.2 | −0.81 | 11.0(4) | −0.68(5) |
| $^{96}$Tc | 0.753 | 40(2) | | | | 11.2 | −1.69 | 11.9 | −1.71 | 11.7(2) | −1.57(2) |
| $^{100}$Tc | 0.711 | 42(2) | 6.765 | 9/2 | 0.0140(12) | 13.17 | −1.35 | 13.89 | −1.38 | 13.70(11) | −1.26(1) |
| $^{102}$Tc | 0.36 | 13(2) | | | | 12.9 | −1.34 | 13.9 | −1.34 | 13.6(2) | −1.23(5) |
| $^{104}$Rh | 0.649 | 51(2) | 6.999 | 1/2 | 0.032(4) | 12.61 | −1.67 | 13.92 | −1.57 | 13.53(17) | −1.48(1) |
| $^{108}$Ag | 0.699 | 46(2) | 7.276 | 1/2 | 0.028(2) | 12.7 | −1.54 | 13.97 | −1.46 | 13.58(8) | −1.37(1) |
| $^{110}$Ag | 0.665 | 47(2) | 6.813 | 1/2 | 0.018(1) | 14.11 | −1.33 | 15.51 | −1.26 | 15.08(6) | −1.18(1) |
| $^{114}$In | 1.112 | 38(2) | 7.274 | 9/2 | 0.0107(8) | 13.81 | −0.77 | 14.66 | −0.79 | 14.40(9) | −0.69(1) |
| $^{116}$In | 0.921 | 36(2) | 6.786 | 9/2 | 0.0095(5) | 14.63 | −0.84 | 15.51 | −0.86 | 15.26(4) | −0.76(2) |
| $^{122}$Sb | 0.474 | 27(2) | 6.807 | 5/2 | 0.013(2) | 13.51 | −1.48 | 14.72 | −1.43 | 14.36(19) | −1.34(1) |
| $^{124}$Sb | 0.484 | 27(2) | 6.467 | 7/2 | 0.024(3) | 12.98 | −1.57 | 14.06 | −1.54 | 13.76(15) | −1.43(1) |
| $^{128}$I | 0.485 | 34(2) | 6.828 | 5/2 | 0.015(2) | 13.21 | −1.66 | 14.42 | −1.6 | 14.07(17) | −1.5(0) |
| $^{130}$I | 0.296 | 26(3) | 6.502 | 7/2 | 0.020(3) | 12.73 | −1.99 | 13.84 | −1.93 | 13.53(15) | −1.82(2) |
| Odd A | | | | | | | | | | | |
| $^{79}$Kr | 1.079 | 32(2) | 8.335 | 0 | 0.25(8) | 9.38 | −1.59 | 10.43 | −1.47 | 10.11(38) | −1.37(7) |
| $^{81}$Kr | 1.351 | 28(2) | 7.874 | 0 | 0.28(8) | 10.22 | −0.94 | 11.35 | −0.86 | 10.99(35) | −0.77(4) |
| $^{85}$Kr | 2.637 | 34(2) | 7.113 | 0 | 4.0(4) | 9.06 | −0.1 | 10.31 | 0.06 | 9.89(10) | 0.11(1) |
| $^{87}$Kr | 2.837 | 35(2) | 5.516 | 0 | 29.0(15) | 8.49 | −0.15 | 10.02 | 0.15 | 9.50(4) | 0.15(3) |
| $^{85}$Sr | 1.85 | 33(2) | 8.527 | 0 | 0.32(12) | 9.91 | −0.63 | 11.01 | −0.53 | 10.65(52) | −0.45(8) |
| $^{87}$Sr | 3.166 | 53(2) | 8.442 | 0 | 2.6(8) | 8.41 | −0.16 | 9.53 | 0.04 | 9.16(41) | 0.09(13) |
| $^{89}$Sr | 3.541 | 32(2) | 6.43 | 0 | 23.7(29) | 8.68 | 0.68 | 10.21 | 0.96 | 9.68(17) | 0.95(1) |
| $^{89}$Y | 3.63 | 26(2) | 11.478 | 4 | 0.106(35)[c] | 8.48 | 0.85 | 9.12 | 0.84 | 8.91(34) | 0.96(6) |
| $^{91}$Zr | 3.053 | 37(2) | 7.261 | 0 | 6.0(14) | 8.92 | 0.17 | 10.22 | 0.37 | 9.77(33) | 0.40(7) |
| $^{93}$Zr | 2.391 | 29(3) | 6.785 | 0 | 3.5(8) | 9.78 | 0.07 | 11.1 | 0.07 | 10.66(29) | 0.12(2) |
| $^{95}$Zr | 2.022 | 14(2) | 6.507 | 0 | 4.0(8) | 10.15 | 0.08 | 11.48 | 0.18 | 11.04(19) | 0.23(5) |
| $^{97}$Zr | 2.058 | 9(1) | 5.619 | 0 | 13(3) | 10.26 | 0.38 | 11.72 | 0.46 | 11.21(30) | 0.51(2) |
| $^{87}$Nb | 1.604 | 12(2) | | | | 8.5 | −0.65 | 10.1 | −0.42 | 9.5(3) | −0.38(3) |

TABLE I. *(Continued.)*

| Nucleus | Fitted level and resonance data | | | | | $I = 0.5I_r$ | | $I = I_r$ | | $I = (0.5–0.75)I_r$ | |
|---|---|---|---|---|---|---|---|---|---|---|---|
| $^{89}$Nb | 1.272 | 9(2) | | | | 8.4 | −0.85 | 10.1 | −0.6 | 9.7(4) | −0.51(6) |
| $^{91}$Nb | 2.66 | 29(4) | | | | 8.9 | −0.07 | 10.2 | 0.12 | 9.3(3) | 0.04(3) |
| $^{89}$Mo | 1.646 | 7(2) | | | | 8.5 | −0.23 | 10.1 | −0.05 | 9.6(3) | 0.01(12) |
| $^{91}$Mo | 2.716 | 29(2) | | | | 8.3 | −0.23 | 9 | 0.09 | 8.95(38) | −0.02(8) |
| $^{93}$Mo | 3.161 | 77(5) | 8.092 | 0 | 2.7(5), 2.17(25)[d] | 8.76 | −0.34 | 9.95 | −0.11 | 9.57(33) | −0.06(8) |
| $^{95}$Mo | 1.743 | 28(2) | 7.379 | 0 | 1.32(18), 0.831(249)[a] | 10.27 | −0.59 | 11.5 | −0.48 | 10.78(63) | −0.48(11) |
| $^{97}$Mo | 1.341 | 33(2) | 6.831 | 0 | 1.05(20), 0.661(198)[a] | 10.67 | −1 | 11.82 | −0.94 | 11.18(71) | −0.91(13) |
| $^{99}$Mo | 1.472 | 49(2) | 5.941 | 0 | 1.0(2) | 11.35 | −0.95 | 12.79 | −0.81 | 12.39(32) | −0.74(4) |
| $^{101}$Mo | 0.626 | 23(2) | 5.411 | 0 | 0.62(10) | 12.14 | −1.36 | 13.62 | −1.25 | 13.14(20) | −1.19(2) |
| $^{91}$Tc | 1.821 | 15(3) | | | | 8.7 | −0.53 | 9.9 | −0.4 | 9.2(4) | −0.40(4) |
| $^{93}$Tc | 2.339 | 21(2) | | | | 8.8 | −0.23 | 9.8 | −0.14 | 9.4(3) | −0.06(1) |
| $^{95}$Tc | 1.433 | 22(2) | | | | 10.1 | −0.82 | 11.4 | −0.69 | 10.9(6) | −0.63(8) |
| $^{101}$Tc | 0.887 | 23(2) | | | | 12.15 | −1.02 | 13.6 | −0.93 | 13.15(40) | −0.85(3) |
| $^{101}$Ru | 1.051 | 28(2) | 6.808 | 0 | 0.345(60) | 11.49 | −1.07 | 12.78 | −0.99 | 12.36(21) | −0.91(1) |
| $^{103}$Ru | 0.954 | 39(2) | 6.239 | 0 | 0.28(5) | 12.31 | −1.2 | 13.71 | −1.1 | 13.27(26) | −1.03(2) |
| $^{105}$Ru | 0.887 | 36(2) | 5.917 | 0 | 0.320(65) | 12.66 | −1.17 | 14.1 | −1.08 | 13.67(30) | −1.00(2) |

[a]Utsunomiya *et al.* [32].
[b]Koehler [33].
[c]Guttormsen *et al.* [34].
[d]RIPL-1 Beijing file [40].

inertia. Thus, the $I/I_r$ ratio is less important only for the level densities with angular momenta around the target-nucleus g.s. spin, as $0^+$ in even-even nuclei. Actually, this premise becomes significant also for low-spin levels but for the odd-odd target nuclei with larger g.s. spin [e.g., see Fig. 1(b) of Ref. []23]. On the other hand, while the corresponding uncertainty for $^{95}$Mo is larger for spin 1/2 levels due to the two distinct $D_0^{exp}$ measured data for this nucleus [31,32], it is significantly lower for high-spin levels.

The above discussion on the spin-cutoff parameter change within an NLD parameter set concerned the excited nucleus $^{95}$Mo due to its role within activation of the well-known 21/2$^-$ isomer $^{93}$Mo$^m$ through $(n, 2n)$ reaction on $^{94}$Mo. The corresponding HF + PE analysis is extended hereafter also to $(p, n)$ reaction on $^{93}$Nb nucleus, with data available until incident energies ∼90 MeV.

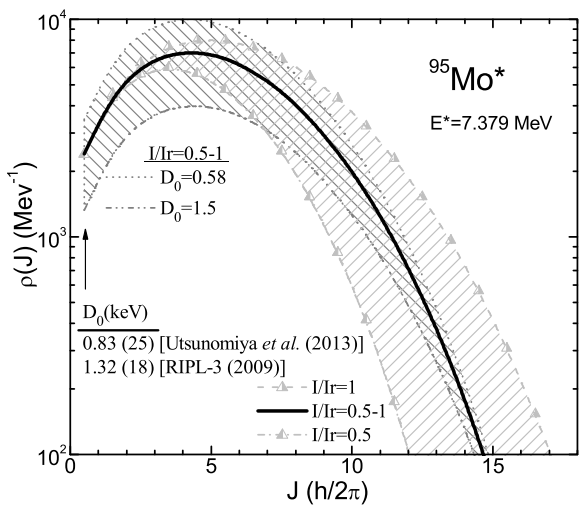

FIG. 2. Spin-dependent level densities of $^{95}$Mo nucleus at *s*-wave average resonance spacing excitation energy, by fit of (1) average $D_0^{exp}$ [31,32], for $\eta$ value of 0.5 (dash-dotted, left-half filled triangles), 1 (dashed, right-half filled triangles), or between 0.5 and 0.75, from g.s. to $S$ [23] (solid), and (2) the lower (short-dashed) and higher (dash-dot-dotted) limits of $D_0^{exp}$, for energy-dependent $\eta$ [23].

## III. SPIN-CUTOFF EFFECTS WITHIN HF + PE MODELING

### A. Models and parameters

In fact, $(n, 2n)$ and $(p, n)$ reactions on $^{94}$Mo and $^{93}$Nb nuclei, respectively, as well as deuteron-induced reactions on Mo isotopes have already been objects of detailed model calculations [1,28,38,45] using consistent input parameter sets involved in the present work, too. Consequently, only main issues are mentioned in the following.

*Statistical model* calculations were carried out within a local approach using an updated version of the HF + PE computer code stapre-h95 [46–49], which includes an additional PE approach within the well-known code stapre [50–52]. In particular, consistent parameter sets of (1) BSFG nuclear level density, (2) transmission coefficients of nucleons, deuterons, α particle, and (3) γ rays were established or validated by means of distinct data. In addition to the NLD parameters detailed in Sec. II, the nucleon [53], α-particle [54], and deuteron [55] optical-model potentials (OMPs), $(p, n)$ reaction cross sections [56], radiative strength functions [57], average *s*-wave radiation widths [31], and $(p, γ)$ reaction cross sections [56] were thus involved. An equidistant binning between 0.1 and 1.3 MeV was used for the excitation energy grid in the present work, from energies close to various reaction thresholds up to above 50 MeV.

*Preequilibrium emission* modeling made use of the geometry-dependent hybrid model [58], generalized by including the angular-momentum and parity conservation [47] and knockout α-particle emission [48] based on a preformation probability [4]. The abovementioned OMP parameters have also been employed for calculation of the intranuclear transition rates as well as within the local-density approximation (LDA) ([58] and references therein). Local-density Fermi energies for various partial waves, corresponding to the central-well Fermi energy value $F = 40$ MeV, have been assumed in this respect. They were determined as a function of orbital angular momentum by an LDA trajectory average using an average imaginary OMP ([58] and Eqs. (58) and (59) of Ref. [49]). Consequently, the first PE nucleon-nucleon interaction predominantly takes place near the nuclear surface [59]. Thus, the particle emission cross-section distribution along the composite system angular momentum [51,52] has included the PE component localized at higher $J$ values (e.g., see Fig. 8 of Ref. [47]). Similarly, the spin-dependent cross sections of the particle emission within stapre basic output [51,52] have concerned both PE and CN components.

At the same time, the revised version of the advanced particle-hole level densities [49,60,61] with a linear energy dependence of the single-particle level density [62] included an advanced pairing correction [63]. A so-called simplified dependence [16] of the PE spin-cutoff parameter by energy and exciton number $n$ has also been used. Actually, the related values are in between $\sigma^2 = 0.16 nA^{2/3}$ initially introduced by Feshbach et al. [14] and the larger value of about $\sigma^2 = 0.28 nA^{2/3}$ obtained by Reffo and Herman [64], particularly for the PE-dominant $n = 2$ component.

*Direct interaction* (DI) collective inelastic-scattering cross sections of nucleons have been obtained by means of the distorted-wave Born approximation (DWBA) method and a local version of the code dwuck4 [65]. Corresponding deformation parameters [66,67] of the first $2^+$ and $3^-$ collective states have been used in this respect. Likewise, the same OMP and level-density parameters have been used in the framework of the HF, PE, and DI models. On the other hand, an appropriate assessment of stripping $(d, n)$ and $(d, p)$ as well as pickup $(d, t)$ and $(d, \alpha)$ direct reactions (DRs) has been obtained [38] through the DWBA method, the appropriate spectroscopic factors, and the code fresco [68].

The *deuteron breakup* (BU), due to deuteron weak binding energy, has also been taken into account by using an empirical parametrization [69–72]. Thus, cross sections were obtained for both elastic breakup, in which the target nucleus remains in its ground state and none of the deuteron constituents interacts with it, and "breakup fusion" (BF) [73], where one of these constituents interacts nonelastically with the target nucleus. The calculated DI cross sections, i.e., BU + DR for deuterons, were then involved for the subsequent decrease of the total-reaction cross sections $\sigma_R$ within HF + PE calculations.

Nevertheless, the excitation functions calculated in this work are also compared with the talys-based evaluated data library TENDL-2023 [74], for an overall excitation function survey.

### B. $\sigma^2$ effects for nucleon-induced reactions

#### 1. $^{94}$Mo$(n, 2n)^{93}$Mo$^m$

A former analysis of the $^{94}$Mo$(n, 2n)^{93}$Mo$^m$ activation using the NLD parameters of the BSFG model using $I/I_r$

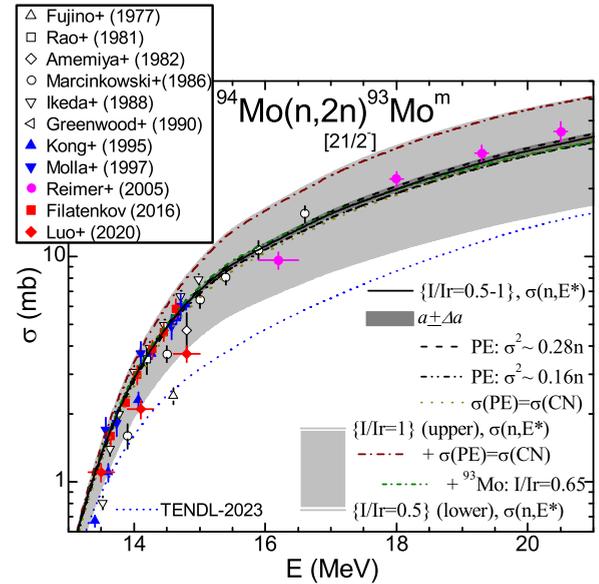

FIG. 3. Comparison of $^{94}$Mo$(n, 2n)^{93}$Mo$^m$ cross sections measured [56], evaluated [74] (dotted), and calculated using (1) the BSFG parameter set (marked by curly brackets) corresponding to the energy-dependent ratio $I/I_r$ [23], and PE spin cutoff either exciton-number $n$- and energy-dependent [16] (solid curve) or (2) like $0.28n$ [64] (dashed) and $0.16n$ [14] (dash-dot-dotted), as well as given by the NLD formula for the CN account (short dotted). The gray uncertainty band corresponds to calculations with the limits of the BSFG parameters $a$ and $\Delta$, while (3) the upper and lower ends of the light-gray band follow use of the parameter sets for constant ratios $I/I_r$ of 1 and 0.5, respectively, and distinct CN and PE spin distributions. Results for using the parameter set $\{I/I_r = 1\}$ but the same CN and PE spin distribution (dash-dotted), and additionally the ratio $I/I_r = 0.65$ for the residual nucleus $^{93}$Mo (short dash-dotted) are also shown.

ratio values of 0.5, 0.75, and 1 at the excitation energies of g.s., neutron binding energy, and 15 MeV, respectively [23], was following the first measurement of corresponding cross sections until an incident energy just above 20 MeV [45]. Actually, an accurate account of the available measured neutron-induced reactions on all stable Mo isotopes in the same energy range was obtained at that time by using a consistent input parameter set. A rather similar set has been involved in the present work, except the fitted data measured in the meantime, in order to get or check the BSFG level-density parameters $a$ and g.s. shifts $\Delta$. However, the reaction mechanism validation was the main goal of the analysis at that date, with particular attention paid to PE formalism. So, while a suitable account of the $^{93}$Mo$^m$ activation by $(n, 2n)$ reaction on $^{94}$Mo nucleus was provided in that instant, no further discussion concerned the NLD issues.

Therefore, an agreement between the measured data and present results obtained by using the BSFG parameters for the energy-dependent $I/I_r$ ratio [23] (latest columns in Table I) and the exciton-number and energy-dependent PE spin cutoff was also expected (Fig. 3). However, in order to illustrate the effects of using various $\eta$ assumptions, these results have been obtained without taking into account also the energy dependence of the level-density parameter $a$ above the neutron binding energy [41–43]. Actually, use of only the NLD parameters in Table I led to an increase of the calculated isomeric cross sections below ≈1% at the upper incident energy.

In these conditions, we note first that, taking into account the limits of the $a$ and $\Delta$ parameters due to the error bars of the $N_d$ and $D_0^{\exp}$ data that were formerly fitted (also in Table I) results in an uncertainty band for the calculated isomeric cross sections of less than 7% (Fig. 3). Comparatively, use of the $a$ and $\Delta$ parameters related to constant $\eta$ values of 1 and 0.5 in Table I, corresponding to a similar fit of the $N_d$ and $D_0^{\exp}$ data, led to an increase and decrease with $\lesssim$50%, respectively, of the calculated results around the upper incident energy. Thus, an uncertainty band larger than 100% at higher energies, with ∼70% around the common 14.5 MeV, is related to the earlier constant $\eta$ values involved in similar fits of the primary data $N_d$ and $D_0^{\exp}$. So, the dominant role of the $I$ assumption for the NLD correctness is obvious, with regard to the parameter accuracy due to the data fitted.

A further point of this analysis concerns the PE spin distribution. Thus, there are shown in Fig. 3 the results of using alternately either the PE spin-cutoff parameters $\sigma^2 = 0.28\,nA^{2/3}$ [64] or $\sigma^2 = 0.16\,nA^{2/3}$ [14]. While the latter is more appropriate for exciton configurations around the most probable exciton number, both of them led to cross-section changes yet close to the uncertainty band related to the data fitted for the BSFG parameter assessment. Actually, a check of also the option $\sigma^2 = 0.24\,nA^{2/3}$ [15], assumed in talys [17], led to results not really distinct from those related to the simplified energy- and $n$-dependent PE spin-cutoff parameter [16]. More important, however, has been the use of the same CN and PE spin distribution, i.e., the spin-cutoff parameter corresponding to the BSFG formula. Lower isomeric cross sections by ∼7% around 14 MeV incident energy and ∼3% around 21 MeV have been obtained in this case. A similar small decrease of $\leqslant$3% has been found for the alternate use of the $a$ and $\Delta$ parameters related to constant $\eta$ value of 1 in Table I (Fig. 3). Nevertheless, it should be noted that the total PE cross section of neutrons on $^{94}$Mo, at the incident energies between 14 and 21 MeV, is from 15% to 28% of $\sigma_R$, respectively.

The final point of this discussion is the use of the BSFG parameter set for $\eta = 1$ and the same CN and PE spin distribution, but additionally a lower $\eta$ value only for the residual nucleus $^{93}$Mo. This is actually the routine way to describe the experimental isomeric cross sections by decreasing the $I$ value. It can be seen that the results corresponding to the value $\eta = 0.65$ in Fig. 3 are rather similar to those obtained in the present work using the BSFG parameter set corresponding to the energy-dependent ratio $I/I_r$ [23], and the exciton-number $n$- and energy-dependent PE spin cutoff [16].

Effects of both the $I/I_r$ and PE spin-distribution assumptions also on the initial occupations preceding secondary emission are shown in Fig. 4. Thus, spin-dependent PE, CN, and PE + CN neutron-emission cross sections of 20 MeV neutrons incident on $^{94}$Mo are calculated by using the BSFG parameter set corresponding to the energy-dependent ratio $I/I_r$ [23] and either the PE spin cutoff [16] or given by the NLD formula for the CN account. They are also compared with the results obtained by using the BSFG parameter set for $I/I_r = 1$ and the same CN and PE spin cutoff. Obviously, the PE spin distribution is peaked at lower spin values since the high-spin states are difficult to make with a simple $1p$-$1h$ configuration [10–12]. This attribute is increased by use of the exciton-number $n$- and energy-dependent PE spin cutoff [16], which reduced even more the higher-spin population provided by the CN spin cutoff. A similar effect is given by the BSFG parameter set for the energy-dependent ratio $I/I_r$ [23] and the same CN and PE spin cutoff.

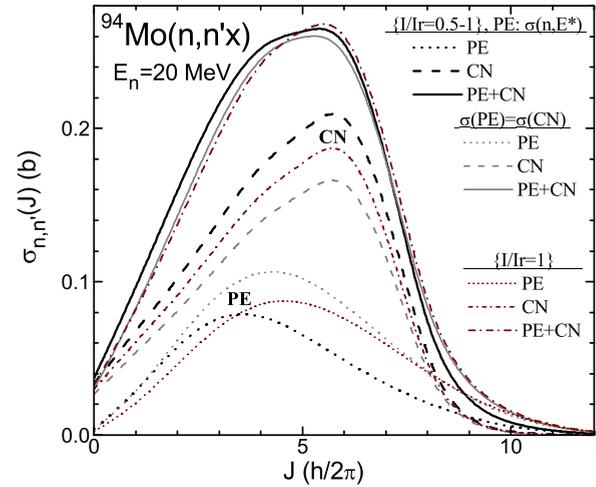

FIG. 4. Comparison of the spin-dependent PE (dotted curves), CN (dashed), and PE + CN (solid) neutron-emission cross sections of 20 MeV neutrons incident on $^{94}$Mo, calculated using (1) the BSFG parameter set corresponding to the energy-dependent ratio $I/I_r$ [23] and either the PE spin cutoff exciton- and energy-dependent [16] (thick curves) or (2) given by NLD formula for the CN account (thin curves), and (3) BSFG parameter set $\{I/I_r = 1\}$ and the same CN and PE spin cutoff (short dotted, short dash-dotted, and dash-dotted curves).

### 2. $^{93}$Nb$(p, n)$$^{93}$Mo$^m$

The rather similar models and parameter sets were previously used for the $(p, n)$ reaction analysis only at lower incident energies ($\leqslant$5 MeV) and particularly for check of the proton OMP parameters. However, unlike the above-discussed $(n, 2n)$ reaction with only two measured datasets available in the latest decade or so, there are in this case more experimental studies with matching cross sections up to the incident energies around 90 MeV [18,75]. This extended energy range thus provides a particular opportunity for a straightforward model validation.

The suitable agreement shown in Fig. 5 with most of the recent data or at least their average has been again achieved by using only the BSFG parameters for the energy-dependent $I/I_r$ ratio [23] and $n$- and energy-dependent PE spin cutoff. Due consideration of also the energy dependence of the level-density parameter $a$ above the neutron binding energy [41–43] corresponds to lower cross sections, from ∼2%–6% in between the incident energies 15 and 50 MeV to 1% at 90 MeV. The latest smaller change may also be related to the PE contribution increase at higher energies, the total PE cross section at proton energies between 15 and 90 MeV, increasing from ∼10% to 89% of $\sigma_R$.





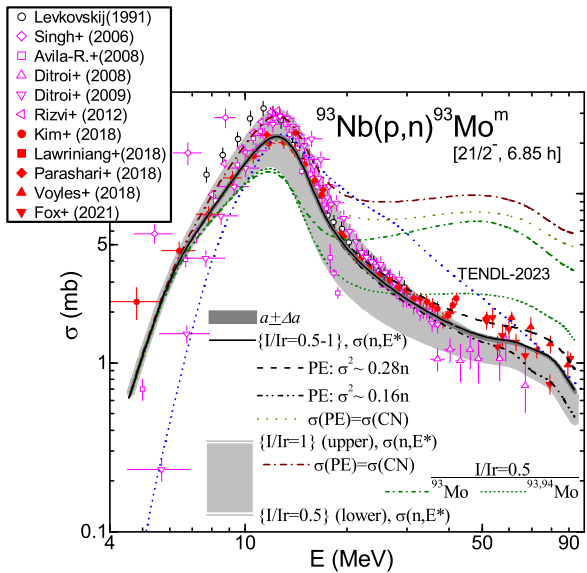

FIG. 5. As Fig. 3 but for $^{93}$Nb$(p, n)^{93}$Mo$^m$ reaction.

Then, an uncertainty band of the calculated isomeric cross sections, with no more than 6% around the incident energy of 15 MeV and ∼3% at lowest and highest energies (the gray band in Fig. 5), has been found corresponding to the limits of the $a$ and $\Delta$ parameters due to the error bars of the $N_d$ and $D_0^{\rm exp}$ data. On the other hand, the use of the $a$ and $\Delta$ parameters related to constant $\eta$ values of 1 and 0.5 in Table I, for a similar fit of the $N_d$ and $D_0^{\rm exp}$ data, led to an increase of ≲40% and a decrease of ≲45%, respectively, of calculated results for incident energies ≲20 MeV [i.e., for rather similar excitation energies as for the above-discussed reaction $(n, 2n)$ on $^{93}$Mo]. Then, the change due to the use of the parameter set with $\eta = 1$ is notably reduced until its vanishing for incident energies above 50 MeV. On the contrary, use of the parameter set with $\eta = 0.5$ corresponds to only a slightly lowering until around 34% at 90 MeV. Therefore, at last, the calculated cross-section uncertainty band due to the various $I$ values still goes from ∼85% at the excitation function maximum to ∼40% at the highest incident energy. The dominant role of the $I$ inference for the NLD correctness, over the precision of the data fitted for the rest of the parameters' assessment, is thus proved also when PE emission prevails.

Concerning the PE spin distribution, the results of using alternately either the PE spin-cutoff parameters $\sigma^2 = 0.28nA^{2/3}$ [64] or $\sigma^2 = 0.16nA^{2/3}$ [14] are shown in Fig. 5, too. Now, there is an increase of the calculated cross sections from ∼5% to 32%, for incident energies from 20 to 90 MeV, for the former case, while a decrease arises in the latter one, from ∼6% to 30% in the same energy range. A check of the option $\sigma^2 = 0.24nA^{2/3}$ [15], assumed in talys [17], was also carried out. Its results, not shown anymore, have been larger within ∼1%–5% than using the simplified energy- and $n$-dependent PE spin-cutoff parameter [16].

Most important has been, however, use of the same CN and PE spin distribution, i.e., the spin-cutoff parameter given by the BSFG parameters. First, lower results by ≲5% have been obtained below the incident energy of 16 MeV in Fig. 5. Then, there is already an increase up to ∼50% around 20 MeV, and

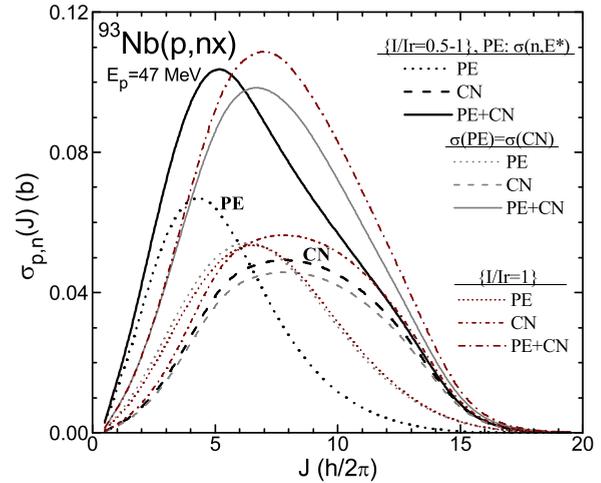

FIG. 6. As Fig. 4 but for 47 MeV protons on $^{93}$Nb.

going to a factor of 6 for energies ⩾60 MeV. A similar trend has been found for the alternate use of the BSFG parameters related to $\eta = 1$ in Table I (Fig. 3), with the same CN and PE spin distribution leading to a final increase by a factor of ∼8 at the highest incident energy. It thus becomes obvious that use of the distinct PE spin distribution is most important at energies above 20 MeV, i.e., much more than the NLD option assumed for the ratio $\eta$.

Use of the BSFG parameter set for $\eta = 1$, the same CN and PE spin distribution, and also an additional value $\eta = 0.5$ only for the residual nucleus $^{93}$Mo led first to results at energies below ∼15 MeV close to use of the NLD parameter set for $\eta = 0.5$. Then, it follows again a cross-section increase, until ≲60% of the former results. In order to have a simplified way of additional changes, the same $\eta = 0.5$ was involved also for the compound nucleus $^{94}$Mo. Again, a similar trend has resulted, close to that using the BSFG parameter set for $\eta = 0.5$ but at energies below ∼20 MeV, and a further increase by ≲25% of the results for the BSFG parameter set with $\eta = 1$.

The latest comment might as well follow the comparison of the spin-dependent PE, CN, and PE + CN neutron-emission cross sections of 47 MeV protons incident on $^{93}$Nb (Fig. 6) similar to that for 40 MeV neutrons in Fig. 4. The larger spin range and PE fraction, due to the higher proton energies, underline the shift to lower $J$ values of the PE component corresponding to the energy-dependent ratio $I/I_r$ [23].

Concluding remarks on both reactions $(n, 2n)$ and $(p, n)$ mentioned above are therefore strongly related to the incident-energy range. First, below 15–20 MeV, there is a 6%–7% uncertainty of the calculated cross section due to either precision of the data fitted for the NLD parameter setup or use of a particular PE spin distribution, or even of the same CN and PE spin-cutoff parameter. It is really less important than 85%–100% incertitude corresponding to the variance of the ratio of the moment of inertia to its rigid-body value. Second, at higher energies, there is a similar uncertainty only with reference to precision of the data fitted for the NLD parameter setup. Third, above 20 MeV, one may find changes from ∼5% to 37% if various options are used for the PE spin distribution [15,16,64], with an additional one (⩽30%) for a former one [14]. Fourth, use of the same CN and PE spin distribution





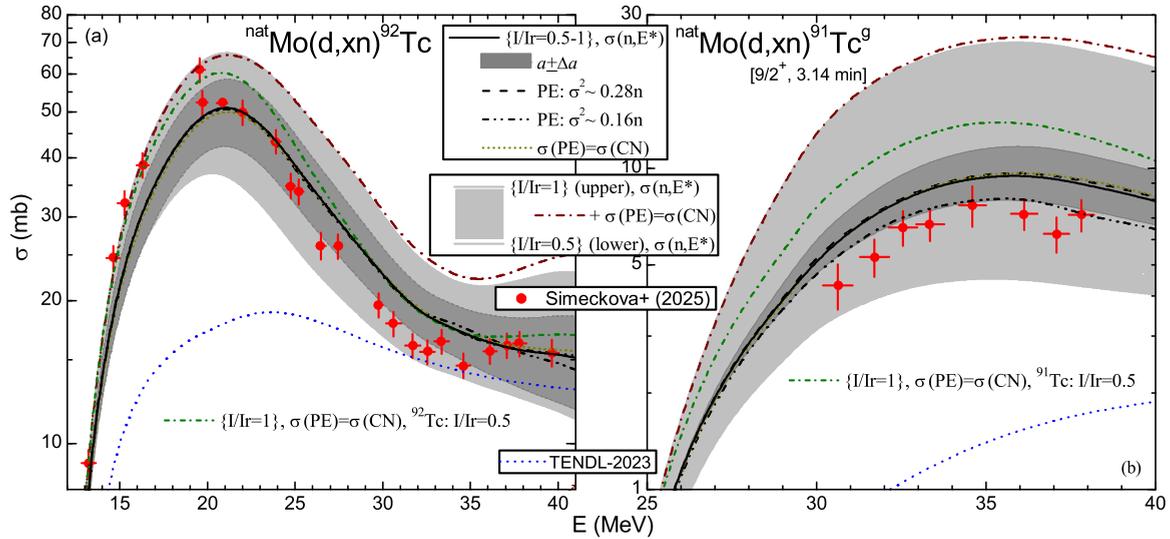

FIG. 7. As Fig. 3 but for activation cross sections of (a) $^{92}$Tc and (b) $^{91}$Tc$^g$ by deuterons on $^{nat}$Mo, and measured data [1] (solid circles).

becomes by far essential, leading to increased results by already 50% around 20 MeV, but factors up to 8 for the BSFG parameter set corresponding to $\eta=1$ values. Fifth, the routine way to describe the experimental isomeric cross sections by using the BSFG parameter set corresponding to the rigid-body value $I_r$ but with half of $I_r$ for the residual nucleus obviously led to decreased cross sections, closer to measured data, at the price of less and less correct NLDs.

### C. $\sigma^2$ effects for deuteron-induced reactions

Previously, only the dominant neutron emission by deuterons incident on $^{nat}$Mo was considered, in order to avoid secondary issues [1]. However, all reaction mechanisms within the complex deuteron interaction with nuclei beyond the usual PE and CN, i.e., the deuteron breakup, stripping, and pickup, were formerly considered also for contributions of specific Mo isotopes [38], making use of the talys default option of the same CN and PE spin distribution. Nevertheless, although $\eta$ values between 0.25 and 1 were taken into account, a suitable data account has been achieved corresponding indeed to $\eta=0.5$ for $^{93}$Tc, $^{93}$Tc$^g$, $^{94}$Tc$^g$, $^{96}$Tc$^m$, and $^{97}$Tc$^m$ activation [1]. At the same time, significant overestimation was found especially at incident energies above $\sim$25 MeV for other states.

A complete analysis of $\sigma^2$ effects is concerned hereafter for deuteron-induced $^{91,92,93}$Tc activation by 40 MeV deuterons incident on $^{nat}$Mo, as well as a proof of their weight even for $^{101}$Tc and $^{101}$Mo activation following the BU and (d, n) and (d, p) mostly direct reactions on $^{100}$Mo nucleus [1].

#### 1. Mo(d, xn)$^{91,92}$Tc

Excitation functions of $^{nat}$Mo(d, xn)$^{92}$Tc and $^{nat}$Mo(d, xn)$^{91}$Tc$^g$ reactions, for incident energies up to 40 MeV, correspond mainly to (d, 2n) and (d, 3n) reactions, respectively, on the molybdenum lightest stable isotope $^{92}$Mo [1]. A significant overestimation by talys default calculations [1] for $^{92}$Tc activation as well as underestimation (Fig. 7) by TENDL evaluation [74] may follow to the excited nucleus $^{94}$Tc former decay through the $N=50$ magic nucleus $^{93}$Tc. On the other hand, their HF + PE analysis is of interest because fewer reaction channels are involved while both an entire residual nucleus and a particular g.s. activation are concerned by this results' discussion.

First, an agreement with data merely within twice their standard deviation is shown hereafter in the whole incident energy range (Fig. 7). The BSFG parameters including energy-dependent $I/I_r$ ratio [23] (latest columns in Table I) and the exciton-number $n$- and energy-dependent PE spin cutoff are used hereof. However, in order to illustrate the effects of various $\eta$ assumptions, the energy dependence of the level-density parameter $a$ above the neutron binding energy [41–43] was not included. The cross-section change due to this simpler approach has been included within the uncertainty band corresponding to the limits of the BSFG parameters $a$ and $\Delta$ related to the error bars of the $N_d$ and $D_0^{exp}$ data formerly fitted (also in Table I). The relative width of this band for $^{nat}$Mo(d, xn)$^{92}$Tc activation goes from $\sim$32% to 42% between the incident energies of 21 and 40 MeV in Fig. 7(a). The larger values at the higher energies follow also the additional contribution of the reaction $^{94}$Mo(d, 4n)$^{92}$Tc, which becomes significant above 34 MeV. The same uncertainty band for $^{91}$Tc$^g$ activation, following entirely the (d, 3n) reaction on $^{92}$Mo isotope, has a 22% width at the energy of 40 MeV [Fig. 7(b)].

Second, the other two BSFG parameter sets corresponding to constant $\eta$ values of 1 and 0.5 but similar fit of the $N_d$ and $D_0^{exp}$ data (Table I) have been additionally used. Consequently, rather similar $\sim$30% increased and decreased results, respectively, for $^{92}$Tc activation are found around 21 MeV, where only (d, 2n) reaction on $^{92}$Mo takes place. On the other hand, results' increase of $\lesssim$50% and decrease of $\lesssim$24%, respectively, were obtained at the upper incident energy in Fig. 7(a). The larger change for $\eta=1$ is again due to the additional contribution of (d, 4n) reaction on $^{94}$Mo. At the same time, use of the two distinct BSFG parameter sets led to $\sim$160% increase and $\sim$49% decrease, respectively, for $^{91}$Tc$^g$ activation at 40 MeV [Fig. 7(b)]. Thus, the dominant role of





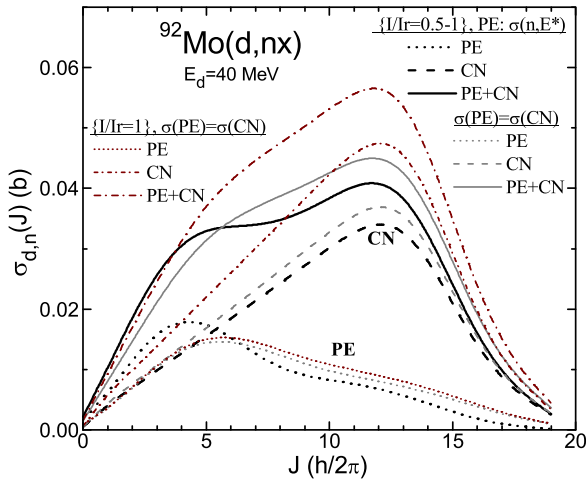

FIG. 8. As Fig. 4 but for 40 MeV deuterons on $^{92}$Mo.

the $I$ assumption, also critical for the NLD correctness, versus the precision of the data fitted for the parameter assessment is obvious.

Third, in addition to the basic use of simplified energy- and $n$-dependent PE spin-cutoff parameter $\sigma^2(n, E^*)$ [16], there are also shown results of using either the PE spin-cutoff parameters $\sigma^2 = 0.28nA^{2/3}$ [64] or $\sigma^2 = 0.16nA^{2/3}$ [14]. While the former is better for the PE-dominant $n=2$ component and the latter is more appropriate for $n$ around the most probable exciton number [16], both of them led to results yet within the uncertainty band related to the limits of the $a$ and $\Delta$ BSFG parameters. Actually, a check of also the option $\sigma^2 = 0.24nA^{2/3}$ [15], assumed in talys [17], led to results not really distinct from those related to the $\sigma^2(n, E^*)$ spin-cutoff parameter [16].

Fourth, use of the same CN and PE spin distribution, i.e., the spin-cutoff parameter corresponding to the BSFG formula, has minor effects too. Cross-section changes below 2% and 4% were found for $^{92}$Tc and $^{91}$Tc$^g$ activations, respectively (Fig. 7). However, the same PE spin distribution replacement corresponds, in the case of alternative use of $a$ and $\Delta$ parameters related to constant $\eta = 1$, to cross-section increase of $\leqslant 8\%$ for both nuclei. This may also be shown by the comparison of the spin-dependent PE, CN, and PE + CN neutron-emission cross sections of 40 MeV deuterons incident on $^{92}$Mo (Fig. 8), much alike that for 47 MeV protons in Fig. 6. It is once again obvious that the shift to lower $J$ values of the PE component corresponds first to the energy-dependent ratio $I/I_r$ [23] and less to the use of distinct CN and PE spin distributions.

Fifth, the ultimate and real object of this analysis has been the use of the BSFG parameter set for $\eta = 1$ and the same CN and PE spin distribution but additionally a value $\eta = 0.5$ for the residual nuclei $^{92,91}$Tc. Actually, it corresponds to the current way to describe the experimental isomeric data by means of a lower $I$ value. A consequent large decrease of $\sim 31\%$ and $\sim 50\%$ provides $^{92}$Tc and $^{91}$Tc$^g$ activations, respectively, in better agreement with measured data at $\lesssim 40$ MeV (Fig. 7). Obviously, a closer accord would be provided by a further decrease of the effective moment of inertia below $I_r/2$ value. The disagreement between the NLD corresponding results and the $N_d$ and $D_0^{\mathrm{exp}}$ data will be, however, even higher.

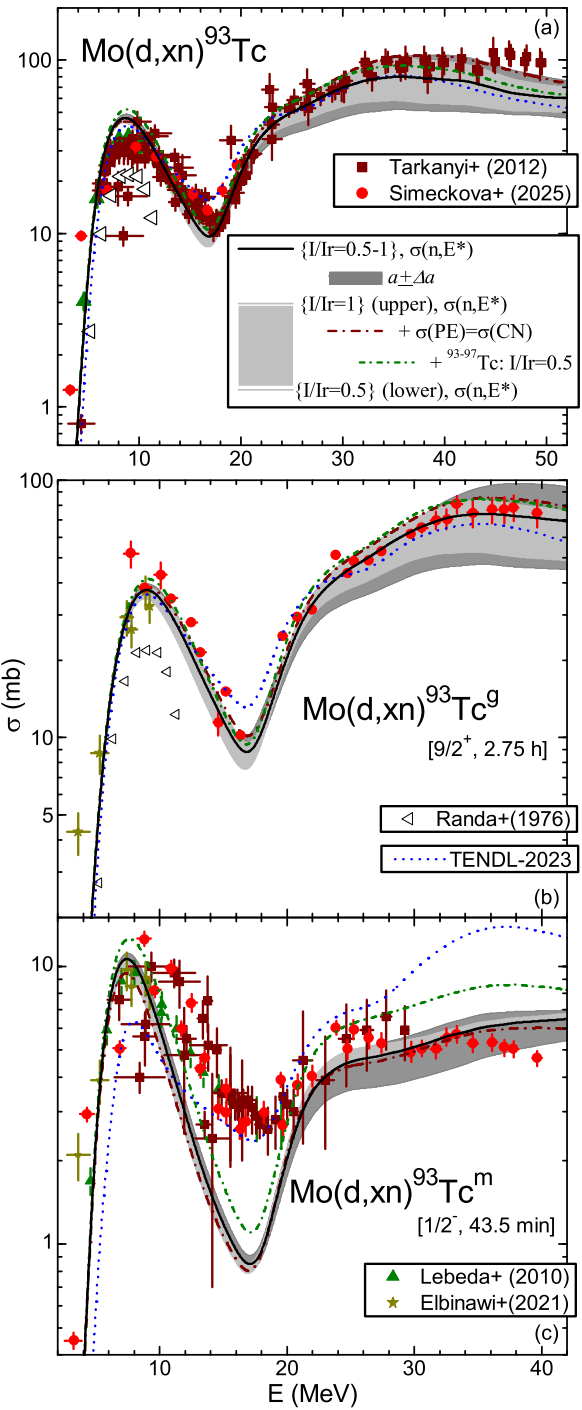

FIG. 9. As Fig. 7 but for (a) $^{93}$Tc, (b) $^{93}$Tc$^g$, and (c) $^{93}$Tc$^m$ activation, and previous measurements [56].

### 2. Mo(d, xn)$^{93}$Tc

The enhanced sequential neutron emission within $^{93}$Tc activation by denterons incident on natural molybdenum is a quite illustrative case. The latest measurement up to 40 MeV [1] is completing most interesting excitation functions for the higher-spin $9/2^+$ g.s., $1/2^-$ metastable state, and total activation cross sections (Fig. 9). Thus, the only $(d, n)$ reaction on molybdenum lightest isotope $^{92}$Mo, up to $\sim 20$ MeV, is followed by the contribution of $(d, 3n)$ reaction on





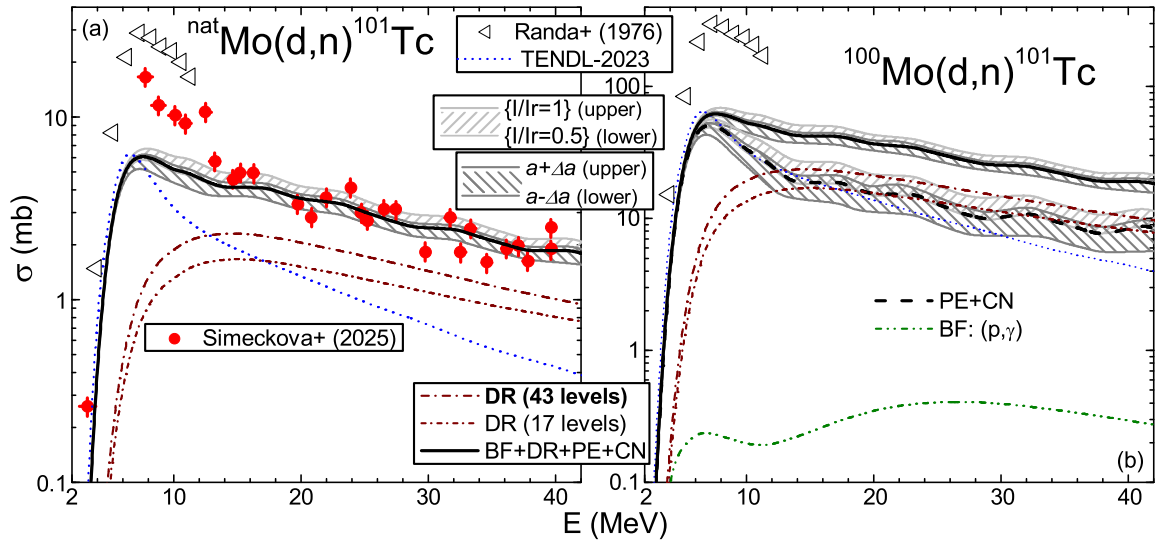

FIG. 10. Comparison of NPI [1] (solid circles) and previously [76] measured, evaluated [74] (dotted curves), and calculated cross sections as in Fig. 3 (solid curves) but for (a) $^{nat}$Mo$(d,n)^{101}$Tc and (b) $^{100}$Mo$(d,n)^{101}$Tc$^+$, as well as PE + CN components (dashed), $(d,n)$ stripping DR using previous spectroscopic factors for either 17 levels [77] (short dash-dotted) or 43 levels (dash-dotted), and BF enhancement (dash-dot-dotted). The gray uncertainty band corresponds to calculations with the limits of the BSFG parameters $a$ and $\Delta$, while the upper and lower ends of the light-gray band follow use of the parameters related to constant $\eta$ values of 1 and 0.5, respectively.

$^{94}$Mo as well as that of $(d,4n)$ reaction on $^{95}$Mo above ∼35 MeV (see Fig. 15 of Ref. [38]). While the results shown in Figs. 9(a)–9(c) have been obtained at once with the calculated $^{91,92}$Tc activation discussed above, several distinct points should be underlined. First, a rather suitable account is obtained for the total and g.s. activations. At the same time, the $1/2^-$ isomer is widely underestimated below 20 MeV, i.e., within the incident energy range where only the $(d,n)$ reaction matters [Fig. 9(c)]. Otherwise, at higher energies the present calculated results are again in agreement with data, which limits the eventual problem to the first-particle emission. However, one should note that former talys calculations showed not only a similar variance at lower energies, but even a larger one above 20 MeV and for any parameter option (Fig. 6 of Ref. [1]).

At the same time, the calculated cross-section uncertainty band due to the use of BSFG parameter sets related to constant $\eta$ values of 1 and 0.5 is larger at incident energies ≲20 MeV than the band corresponding to the limits of the $a$ and $\Delta$ parameters for $\eta$ between 0.5 and 1 [23]. Then, effects of the BSFG parameter limits for more residual nuclei at higher energies increase all together their uncertainty band, as, e.g., in Fig. 7(a) above 35 MeV. Thus, the relative width of the corresponding band is ∼72% at 40 MeV for both $^{93}$Tc and $^{93}$Tc$^g$ activations, but only 54% and 40% for the former uncertainty band in Figs. 9 (a) and 9(b), respectively.

Nevertheless, this uncertainty band related to constant $\eta$ values 1 and 0.5 is only ≲4% in the case of $^{93}$Tc$^m$ activation [Figs. 9(c)], i.e., lower by at least an order of magnitude with reference to $^{93}$Tc$^g$ and $^{93}$Tc total activation. There is also a decrease but only to ∼27% for the uncertainty band related to the limits of the BSFG parameters $a$ and $\Delta$. Thus, one may presume that the model underestimation is related to either the spin distribution of the $N = 50$ magic nucleus $^{93}$Tc or its

particular structure and decay scheme still available. On the other hand, a further attention should be paid to the BU cross-section consideration for the subsequent decrease of the total-reaction cross sections $\sigma_R$ within HF + PE calculations. Thus, the eventual nuclear surface localization of the deuteron BU, similarly to DR and PE [59], is not taken into account also for depletion of the higher partial-wave occupation. Further work should concern schematic decomposition of the total-reaction cross section with respect to the impact parameter, into CN, DR, and BU cross sections (e.g., see Fig. 1.8 of Ref. [78]).

### 3. Mo$(d,n)^{101}$Tc

The deuteron interaction with only the neutron-richest stable isotope $^{100}$Mo is involved within $^{101}$Tc residual nucleus activation by deuterons incident on $^{nat}$Mo. This analysis concerns the $^{101}$Tc excitation function reported for the first time for a natural molybdenum target [1,38] [Fig. 10(a)], in addition to the previously measured $^{100}$Mo$(d,n)^{101}$Tc activation cross sections [76] [Fig. 10(b)].

While the inelastic breakup enhancement through $(p,\gamma)$ reaction on $^{100}$Mo remains much lower in the whole energy range, the calculated cross sections for $^{100}$Mo$(d,n)^{101}$Tc reaction [Fig. 10(b)] proved formerly [38] a significant DR stripping $(d,n)$ component only for incident energies above 20 MeV. Because an evident underestimation of the measured cross-section maxima was already noted [1,38], an additional stripping contribution has been found in the meantime corresponding to the use of spectroscopic factors for a number of $^{101}$Tc levels increased from 17 [77] to 43. The increase of the resultant DR component goes from ∼40%, at its maximum around the incident energy of 15 MeV, to ∼25% at 40 MeV. It led to DR prevalence versus the statistical PE + CN from even lower energies.





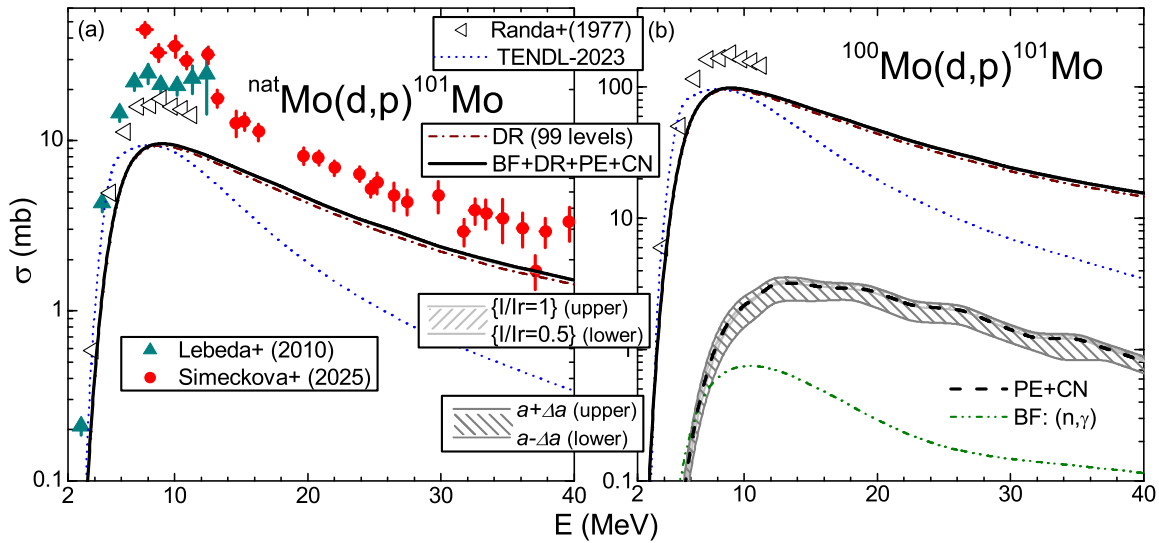

FIG. 11. As Fig. 10 but for $^{101}$Mo activation NPI data [2] and previous measurements [76].

However, the accuracy of the statistical component for the total cross section of the $(d,n)$ reaction on $^{100}$Mo is also of interest. First, taking into account the limits of the $a$ and $\Delta$ parameters due to the error bars of the $N_d$ and $D_0^{\rm exp}$ data results in an uncertainty band for the calculated total cross sections of around 23% up to the incident energy of 40 MeV (Fig. 10). Comparatively, the use of the $a$ and $\Delta$ parameters related to constant $\eta$ values of 1 and 0.5 in Table I, corresponding to a similar fit of the $N_d$ and $D_0^{\rm exp}$ data, led to an uncertainty band of $\sim$15% around the same upper incident energy. It should be noted, however, that the former band is enlarged by considering the energy dependence of the level-density parameter $a$ above the neutron binding energy [41–43], which leads to a minor cross-section decrease.

Nevertheless, there are two particular issues. First, it seems that rather similar effects on the total cross sections are due to the precision of the data fitted for the NLD parameter assessment as well as the $I$ assumption. Second, they are close to the change of the DR component due to the increase of the levels for which further spectroscopic factors have been taken into account. Moreover, all of them are comparable to the average spreading of the newly measured $^{\rm nat}$Mo$(d,n)$$^{101}$Tc cross sections shown in Fig. 10(a) for incident energies above 12 MeV, which are rather well described within this energy range.

There is yet a notable underestimation of the measured cross-section maximum at energies below $\sim$12 MeV. It is more probably due to DR component miscount, eventually following still less appropriate spectroscopic factors for the $(d,n)$ reaction, while the NLD effects are somewhat lower. This issue is supported by the apparent data underestimation by the TENDL evaluation, which obviously neglects the stripping $(d,n)$ contribution especially above 12 MeV [Fig. 10(a)].

#### 4. Mo$(d,p)^{101}$Mo

Similar to the case of $^{101}$Tc residual nucleus, the activation of $^{101}$Mo by a $(d,p)$ reaction follows the deuteron interaction with only the neutron-richest $^{100}$Mo stable isotope. It is why the analysis of deuteron-induced reactions even on a natural Mo target may provide a valuable check of the model calculations having to take care of the BU, DR, and statistical PE + CN mechanisms. However, concerning the aim of this work related to NLD effects on calculated cross sections, its benefit is much lower due to the great DR component. The former DWBA assessment [38] based on the use of the spectroscopic factors for 99 states of $^{101}$Mo nucleus up to the excitation energy of 3.268 MeV, and yet associated with only $l = 0$ and 2 [77], led to a DR weighting from 98% to 94% for deuteron energies between 7 and 40 MeV in Fig. 11(b). The obvious underestimation by a factor $\sim$2 of the measured $(d,p)$ cross sections for both the natural Mo and $^{101}$Mo isotopes (Fig. 11) can therefore be related only to questions of spectroscopic data already open to discussion [77].

Nevertheless, for the sake of discussion, calculated total cross-section uncertainty band due to limits of the $a$ and $\Delta$ parameters following the error bars of the formerly fitted $N_d$ and $D_0^{\rm exp}$ data has a width between $\sim$56% at the deuteron energy of 7 MeV and 27% at 40 MeV. On the other hand, the use of the $a$ and $\Delta$ parameters related to constant $\eta$ values of 1 and 0.5, for a similar fit of the $N_d$ and $D_0^{\rm exp}$ data, led to an uncertainty band of $\leqslant$15% and thus less visible in Fig. 11(b). Obviously, these uncertainty bands are already not distinct after addition of the DR contribution to the total $(d,p)$ reaction for either $^{101}$Mo isotope or natural Mo target in Fig. 11(a).

A final remark following this concurrent discussion under similar conditions of the $(d,n)$ and $(d,p)$ on natural Mo concerns the DR role that is by far dominant only for the latter. Otherwise, it is alike the statistical one for the former. In the latest case, the DR uncertainties due to the available spectroscopic data basis may have comparable effects for the calculated cross sections as the accuracy of the data fitted to obtain the NLD parameters. Thus, the suitable assumption on the spin-cutoff parameter is of interest also for the assessment of total-reaction cross sections as well as for similar DR and statistical components.

## IV. CONCLUSIONS

Various NLD parameter sets are obtained by fit of low-lying discrete level numbers and average *s*-wave resonance spacings, for distinct $I$ values between full and half of its rigid-body value. Parameter limits due to the fitted data error bars are also obtained, corresponding to each $I$ option. Replacement of the $I_r$ value with half of it, within a given parameter set, is shown to demand a change of the related NLD parameter $a$ significantly beyond its fitted limits, to have yet the NLD corresponding to the originally fitted data.

A first discussion of the spin-cutoff parameter change within an NLD parameter set and the use of either the same or distinct CN and PE spin distributions involved the activation of the well-known $21/2^-$ isomer $^{93}$Mo$^m$ by $(n,2n)$ and $(p,n)$ reactions on $^{94}$Mo and $^{93}$Nb nuclei, respectively. Then, complete analysis of these effects has concerned $^{91,92,93}$Tc activation by deuterons incident on $^{nat}$Mo, implying still fewer Mo stable isotopes. The dominant role of the $I$ assumption, also critical for the NLD correctness, versus the accuracy of the data fitted for the parameter assessment is obvious. On the other hand, various PE spin-cutoff parameters led to cross-section changes yet within the uncertainty band related to the limits of the $a$ and $\Delta$ parameters.

Nevertheless, use of $\eta = 0.5$ value only for the residual nucleus, in addition to the BSFG parameter set for $\eta = 1$ and the same CN and PE spin distribution (i.e., the current way to describe the experimental isomeric cross sections by decreasing the $I$ value), provides activation results in better agreement with the measured data on the basis of inexact NLDs. Obviously, a further decrease of $I$ would be even better, at the price of less and less correct NLDs. Finally, the dominant role of the $I$ assumption for NLD correctness within consequent reaction calculations proves the value of the direct method of Weigmann *et al.* [22] to endorse the moment of inertia handling the NLD spin distribution. Further measurements not only of isomeric cross sections but also of average resonance spacings of *s*-wave neutrons and protons, corresponding to different spins of the same nucleus, are thus highly demanded.

## ACKNOWLEDGMENTS

The authors would like to acknowledge the professional and dedicated assistance from GANIL staff, and to thank the anonymous referee for helpful comments on the deuteron-breakup surface localization. This work has been partly supported by MEYS, Czech Republic, under the project SPIRAL2-CZ, as well as the Romanian Nucleu Project No. PN 23 21 01 02, and carried within the framework of the EUROfusion Consortium, funded by the European Union via the Euratom Research and Training Programme (Grant Agreement No. 101052200—EUROfusion). Views and opinions expressed are, however, those of the author(s) only and do not necessarily reflect those of the European Union or the European Commission. Neither the European Union nor the European Commission can be held responsible for them.

## DATA AVAILABILITY

The data that support the findings of this article, beyond the content of Table I, are openly available [19].